# Probing the microscopic structure of bound states in quantum point contacts

Y. Yoon[1], L. Mourokh[2,3], T. Morimoto[4], N. Aoki[5], Y. Ochiai[4,5], J. L. Reno[6], and J. P. Bird[1]

1: Department of Electrical Engineering, University at Buffalo, the State University of New York, Buffalo, NY 14260-1920, USA

2: Department of Physics, Queens College of CUNY, 65-30 Kissena Blvd., Flushing, NY 11367, USA

3: Quantum Cat Analytics, Brooklyn, NY 11204, USA

4: Graduate School of Science and Technology, Chiba University, 1-33 Yayoi-cho, Inage-ku, Chiba 263-8522, Japan

5: Department of Electronics and Mechanical Engineering, Chiba University, 1-33 Yayoi-cho, Inage-ku, Chiba 263-8522, Japan

6: CINT Science Department, Sandia National Laboratories, P.O. Box 5800, Albuquerque, NM 87185-1303

Using an approach that allows us to probe the electronic structure of strongly pinched-off quantum point contacts (QPCs), we provide evidence for the formation of self-consistently realized bound states (BSs) in these structures. Our approach exploits the resonant interaction between closely-coupled QPCs, and demonstrates that the BSs may give rise to a robust confinement of single spins, which show clear Zeeman splitting in a magnetic field.

There has long been great interest in the phenomena arising from the interactions of carriers in nanostructures, such as nanowires and quantum dots. In few-electron quantum dots, for example, the exchange interaction causes a filling of electron spins in a similar manner to Hund's rule [1]. The work described here, on the other hand, is motivated by recent interest in correlated electron transport in quantum point contacts (QPCs) [2-17], which are one-dimensional conductors whose carrier density may be tuned by means of their gate voltage. There have been many suggestions that spin degeneracy may be spontaneously broken in QPCs, under conditions where the density (and conductance) is about to vanish. Most studies of this problem have focused on the so-called 0.7 feature [2], an anomalous plateau-like structure in the conductance that occurs for the range of gate voltage where the QPC is *partially* transmitting. It has recently been suggested, however, that a precursor to this regime should involve binding of single spins to QPCs, for stronger gate confinement where their conductance is *quenched* [17]. While this regime is inaccessible to experiments on single QPCs, in this Letter we provide evidence for spin binding by exploiting the resonant interaction between a bound spin on one QPC and a second that serves as a detector. In this way, we infer important microscopic information on the naturally-formed bound state (BS) that confines the single spin, including its effective confinement and spin structure. Our finding that QPCs may serve as a naturally-formed, electrically addressable, single-spin system could have future applications to spintronics and spin-based quantum computing.

Previously, we showed that a ("detector") QPC exhibits a resonance when it is coupled to another ("swept") QPC that is pinching-off [18,19]. Motivated by the idea that the QPC forms a quantum-dot like potential near pinch-off [14,16], we developed a model [20] relating this resonance to the formation of a BS in the swept QPC. As the swept-QPC gate voltage ($V_g$) is made more negative, the BS is driven up in energy, until, when it aligns with the Fermi energy ($E_F$), the detector-BS interaction produces a Fano-type resonance in the detector conductance ($G_d$) [20]. Only a single resonance is observed in $G_d$, which we attribute [20] to the existence of a large Coulomb energy [16] that blocks the population of the BS by more than a single electron. While this phenomenological model therefore captures the idea of *single-electron* binding, it *cannot* predict the microscopic structure of the BS *per se*. As we now discuss, this issue must instead be addressed through experiment.

Split-gate QPCs were formed in GaAs/AlGaAs quantum wells (35-nm wide) with a two-dimensional electron gas (2DEG) 200-nm below the surface. At 4.2 K, the 2DEG density, mobility, and mean-free path were $2.3 \times 10^{11}$ cm$^{-2}$, $4 \times 10^6$ cm$^2$/Vs, and 32 µm, respectively. Conductance was measured by low-frequency lockin detection (fixed excitation of 30 µV), in a cryostat with a base temperature of 4.2 K. Our devices (Fig. 1) have eight independent gates and, grounding half of these at any one time, we use the remaining four to realize swept and detector QPCs. Figure 1 shows the results of conductance measurements using different QPCs. Black curves in each panel show the variation of the swept-QPC conductance, as a function of $V_g$ and with fixed voltage applied to the detector-QPC gates. Red curves, however, show the variation of the detector conductance for exactly the same gate conditions. Each panel shows a peak in $G_d$ that occurs *just after* the swept QPC pinches off. (Note that at 4.2 K, the 1D conductance quantization is washed out in our devices and only the 0.7 feature survives [2,21]). The isolated peak is consistent with our theory [20], which ascribes the resonance to the occupation of a self-consistently formed BS by just a single electron. The similarity of the measurements in Fig. 1 rules out random impurities, or *unintentional* quantum-dot formation, as the source of the BS, and points instead to a systematic phenomenon involving the swept QPC. Indeed, the behavior here was reproduced in measurements made over fifteen months, on six different thermal cycles, and was also found previously in studies of a very different gate geometry [18].

Figure 2(a) shows that the detector peak persists weakly even at 35 K, behavior that is reproduced quantitatively for other QPC combinations. This suggests that the effective confinement associated with the swept-QPC BS is of order a few meV, although this should be considered as a lower bound since other mechanisms could cause the quenching. (For example, electron dephasing in the connecting region of 2DEG.) The peak also shifts to more-negative $V_g$ with increasing temperature, suggesting stronger gate confinement is needed to form the self-consistent BS at higher temperatures. Figure 2(b) shows the temperature dependence of the peak amplitude ($\Delta$) and its full-width at half maximum ($\Gamma_{HM}$). $\Delta$ saturates below ~8 K, indicating full development of the resonance, but decreases by more than an order of magnitude on reaching 35 K. $\Gamma_{HM}$ is almost constant over this entire range (increasing by ~20%), however, indicating the peak does *not* arise from any internal interaction *within* the detector (in this situation, thermal smearing in the reservoirs should yield a broadening proportional to temperature). The observed behavior is instead consistent with Ref. 20, which attributes the peak to the resonant interaction between a plane wave in the detector and the BS. The plane wave is formed from the superposition of degenerate quantum states at $E_F$, associated with the occupied 1D subbands in the detector. The coupling of both QPCs to their reservoirs broadens the states involved in this interaction, although the magnitude of this broadening should be much larger for the detector than the BS. Since $\Gamma_{HM}$ should reflect the total level broadening, it is therefore not surprising that it varies only weakly in experiment, somewhat reminiscent of the behavior found for the tunneling resonances of coupled quantum dots [22].

The idea that $\Gamma_{HM}$ reflects the overlap of levels in the swept and detector QPCs is consistent with the evolution of the detector peak with $G_d$ (Fig. 3). When $G_d > G_o$, the occupied subbands of the detector are almost fully transmitted and we do not expect their broadening to vary significantly with $G_d$. This is what we find in experiment, for which both $\Delta$ and $\Gamma_{HM}$ are almost independent of $G_d$ in this regime (Fig. 3(b)). For $G_d < G_o$, however, only the lowest subband is occupied in the detector and its transmission is strongly dependent on $V_g$. In this limit we expect the broadening of this level to decrease with $G_d$, causing a reduction of $\Gamma_{HM}$. Such behavior is clearly seen in experiment Fig. 3(b), further supporting the idea that the detector peak arises from the interaction of broadened states in the detector and a much sharper one in the swept QPC.

An in-plane magnetic field should cause the BS to develop Zeeman splitting, forming two levels for electron occupation [16]. The one-electron ground state will then correspond to the lower Zeeman branch [23], which shifts to lower energy with increasing *B*. Accompanying this, the detector peak should therefore shift to more negative $V_g$, since stronger gate confinement is needed to bring this branch into resonance with $E_F$. Fig. 4(a) shows results consistent with this, with a linear shift of the detector peak to more-negative $V_g$ as $B$ is increased (black symbols). As we show in Fig. 4(b), this shift is accompanied by the development of a very weak shoulder (see arrows) on the high-energy (less-negative $V_g$) side of the peak. (Note the decrease of $G_d$ with increasing $B$ in Fig. 4(b). While a more detailed analysis will be presented elsewhere, we attribute this to the diamagnetic shift of the 2DEG subbands [3]). The development of the shoulder is shown more clearly in Fig. 4(b) as an additional feature that appears in $|dG_d(V_g)/dV_g|$. The $V_g$ position of the shoulder is plotted with red symbols in Fig. 4(a), and shows a linear shift with opposite slope to the main peak. The two data sets in Fig. 4(a) extrapolate to a *non-zero* separation at $B = 0$, with $\Delta V_g(B = 0)$ ~33 mV. By assuming a g-factor of 0.4, we can convert the slopes in Fig. 4(a) to an energy change and relate $\Delta V_g(B = 0)$ to an energy splitting of ~0.8 meV. Enhanced g-factors (~1.5) have been reported [2,3] for QPCs near pinch-off, however, so the value may actually be much larger than this (consistent with the washout temperature of the detector peak).

The Zeeman shift of the main peak in Fig. 4(a) is consistent with single-electron occupation of the BS. As for the high-energy shoulder, this might arise from the population of the *upper* Zeeman branch by a *second* electron. In this case, the energy splitting at $B = 0$ would correspond to the Coulomb energy for adding the electron [16]. An alternative interpretation, however, relates these features to the ground and excited states of a *single* electron on the BS [23]. While the lower Zeeman branch is the most favored for occupation, there is a much smaller (thermal) probability that the electron may occupy the *upper* branch, while leaving the lower one empty. Unlike the ground state, the energy of this excited state *increases* with $B$ and its low probability of occupation means that it should give rise to only a *weak* feature in the conductance, just as we find in experiment [23]. (The population of the BS for the two situations is shown schematically in the insets to Fig. 4(a).) This picture seems to account quantitatively for the very different amplitudes of the main resonance and the shoulder, and therefore suggests that the non-zero energy splitting in Fig. 4(a) in fact arises from a spontaneous spin polarization of the BS at $B = 0$.

In conclusion, spin binding in QPCs has been detected as a resonance in the detector QPC that *systematically* occurs for *stronger* gate confinement than the 0.7 feature. This suggests that this resonance and the 0.7 feature are *separate* manifestations of the common phenomenon of spin polarization in QPCs. While we detect the localized spin when it is strongly bound to the pinched-off QPC, the 0.7 feature is seen when the gate confinement is weakened and the QPC becomes partially transmitting. Our results should therefore provide a valuable alternative starting point for understanding the origins of the 0.7 feature.

*Acknowledgement*: The authors gratefully acknowledge discussions with K.-F. Berggren, Y. Meir, and M. Pepper. This work was supported by the Department of Energy and was performed, in part, at the Center for Integrated Nanotechnologies, a U.S. DoE, Office of Basic Energy Sciences nanoscale science research center. Sandia National Laboratories is a multi-program laboratory operated by Sandia Corporation, a Lockheed-Martin Company, for the U. S. Department of Energy under Contract No. DE-AC04-94AL85000.

**FIGURE LEGENDS**

**Figure 1.**   Electron micrograph at top shows device with gate/Ohmic-contact numbering schemes. Lower panels: resonant QPC interaction for different QPC pairs. Black curves: variation of swept-QPC conductance with $V_g$. Red curves: $G_d(V_g)$, with fixed voltage applied to detector gates. Arrows indicate $0.7G_o$, and detector (red) and swept (black) QPCs are indicated in the panel insets. Dotted lines show grounded gates. See [18] for details.

**Figure 2.**   (a) Temperature-dependent evolution of the detector peak. Detector QPC: $G_7$ & $G_8$. Swept QPC: $G_1$ & $G_2$. (b) Temperature-dependent variation of the peak amplitude ($\Delta$) and the full-width at half maximum ($\Gamma_{HM}$).

**Figure 3.**   (a) $V_g$-dependent evolution of the detector peak. Detector QPC: $G_7$ & $G_8$. Swept QPC: $G_1$ & $G_2$. (b) $G_d$-dependent variation of the peak amplitude ($\Delta$) and the full-width at half maximum ($\Gamma_{HM}$).

**Figure 4.**   (a) $B$ dependence of $V_g$ position of main peak (black circles) and weak shoulder (red circles). Dotted lines are linear fits. Detector: $G_7$ & $G_8$. Swept QPC: $G_1$ & $G_2$. Insets: population of the upper and lower Zeeman branches of the BS, corresponding to red and black data sets. BS is shown with weaker confinement in the red schematic, since it corresponds to less-negative $V_g$ than the main peak. Dotted lines: $E_F$ in the 2DEG. (b) $G_d(V_g)$ and $|dG_d/dV_g|$ at several $B$. Data offsets are indicated.

**FIGURE 1**

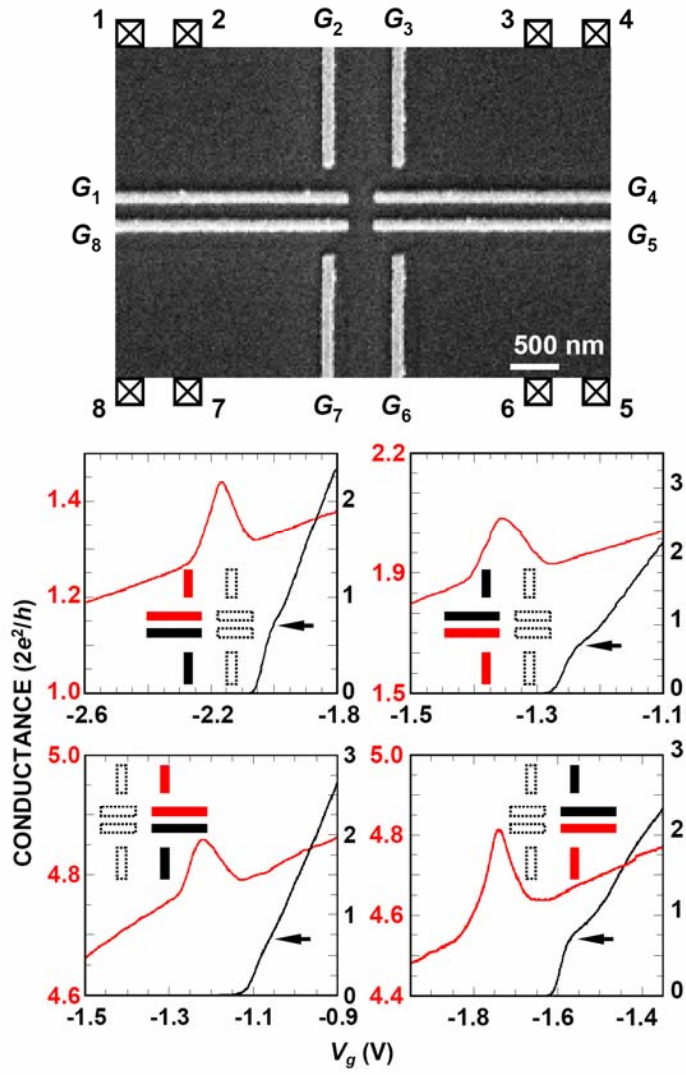

**FIGURE 2**

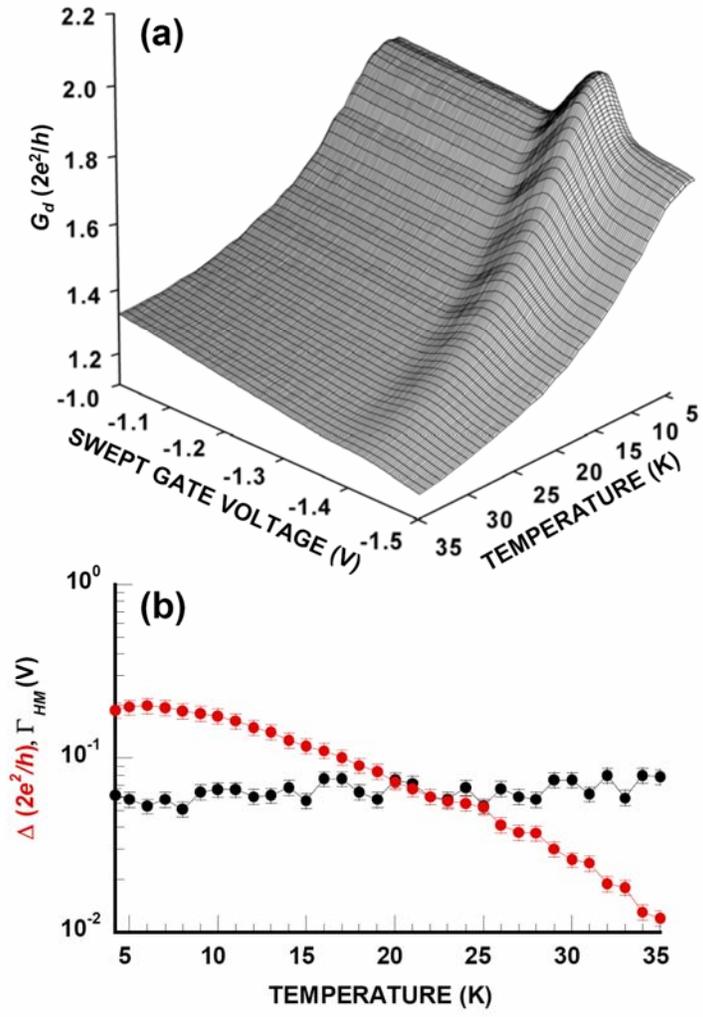

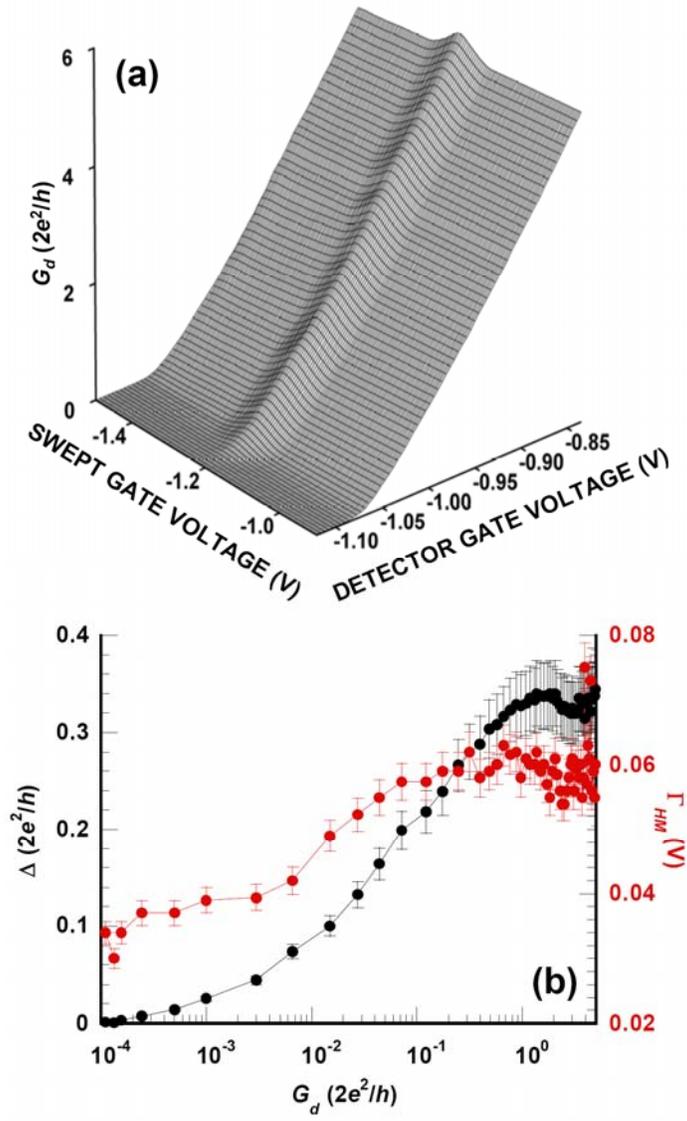

**FIGURE 4**

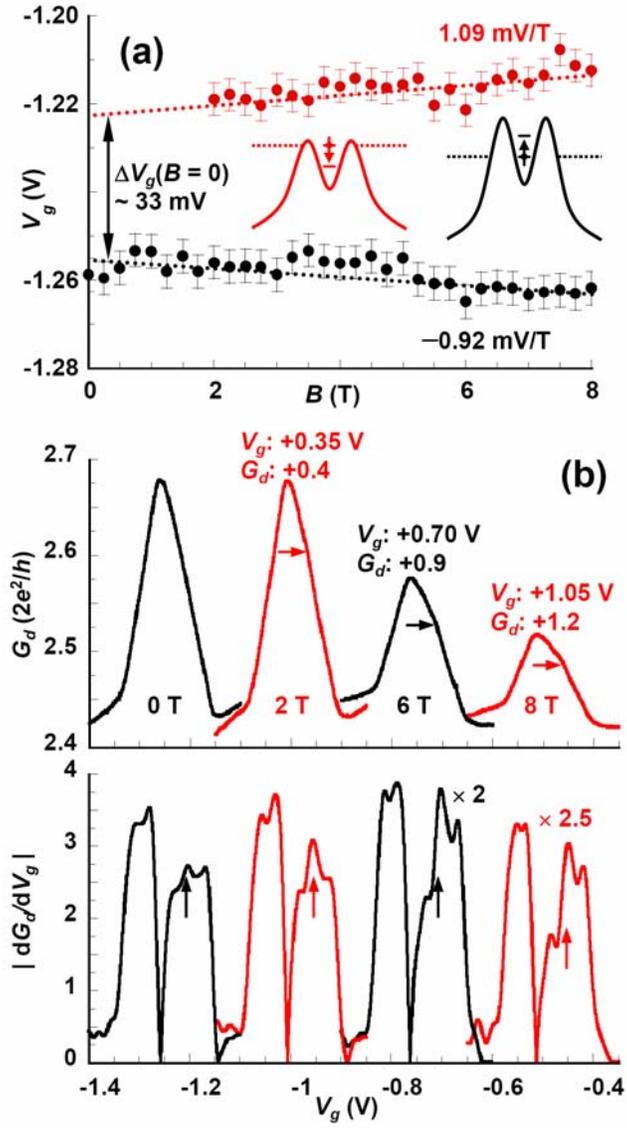